\PassOptionsToPackage{url}{hyphens}
\PassOptionsToPackage{hyperref}{hidelinks=true,breaklinks=true}
\documentclass[pageno,nohyperref, anonymous]{jpaper}

\usepackage{adjustbox}
\usepackage{siunitx}
\usepackage{multirow}
\usepackage[linesnumbered,ruled,vlined]{algorithm2e}
\usepackage{inconsolata}

\usepackage{fancyhdr}
\usepackage{amssymb}
\usepackage[normalem]{ulem}
\usepackage[sort,nocompress]{cite}
\usepackage{authblk}
\usepackage{xstring}
\usepackage[usenames,dvipsnames]{xcolor}
\usepackage{enumitem}
\usepackage[italic]{mathastext}
\usepackage[hang,bottom]{footmisc}
\usepackage[all]{nowidow}
\usepackage{tikz}
\usepackage[hidelinks,breaklinks=true]{hyperref}
\usepackage{MnSymbol,wasysym}

\newif\ifcameraready
\camerareadytrue

\newif\ifonlineversion
\onlineversionfalse

\DeclareSIUnit{\nothing}{\relax}

\begin{document}

\title{Block-SSD: A New Block-Based Blocking SSD Architecture}

\renewcommand\Authfont{\aufnt}
\renewcommand\Affilfont{\affaddr}
\newcommand{\tsc}[1]{\textsuperscript{#1}} 
\newcommand{\affilASS}{$^\dagger$}
\newcommand{\affilASL}{$^\ddagger$$^\ddagger$$^\ddagger$}

\newcommand{\circled}[1]{%
  \tikz[baseline=(char.base)]{
  \node[shape=circle,draw,inner sep=0.5pt,fill=black,text=white,font=\small\bfseries] (char) {#1};}%
}

\ifcameraready
\author{
    Ryan Wong\affilASL\qquad
    Arjun Tyagi\affilASS\qquad
    Sungjun Cho\affilASS\qquad
    Pratik Sampat\affilASS\qquad
    Yiqiu Sun\affilASS\qquad
}
\affil{
    \affilASS Arcana Spirit Squad \quad
    \affilASL Arcana High Leadership Council\quad
   \vspace{-14pt}
}
\else
\author{
    Ryan Wong\affilASL\qquad
    Arjun Tyagi\affilASS\qquad
    Sungjun Cho\affilASS\qquad\\
    Pratik Sampat\affilASS\qquad
    Yiqiu Sun\affilASS\qquad
}
\affil{
    \affilASS Arcana Spirit Squad \quad
    \affilASL Arcana High Leadership Council\quad
   \vspace{-14pt}
}
\fi

\date{}

\maketitle

\ifcameraready
\setstretch{0.86}
\else
\setstretch{0.86}
\fi
\renewcommand{\footnotelayout}{\setstretch{0.87}}
\captionsetup[figure]{font={sf,bf,stretch=0.9}}
\newcommand\blfootnote[1]{%
  \begingroup
  \renewcommand\thefootnote{}\footnote{#1}%
  \addtocounter{footnote}{-1}%
  \endgroup
}

\thispagestyle{fancy}
\fancyhf{}

\section{Introduction}
Computer science and related fields (e.g., computer engineering, computer hardware engineering, electrical engineering, electrical and computer engineering, computer systems engineering) often draw inspiration from other fields, areas, and the real world in order to describe topics in their area.
One cross-domain example is the idea of a \textbf{block}.
The idea of blocks comes in many flavors, including software (e.g., process control blocks, file system blocks, data blocks, basic blocks, blocking statements, blocking processes, blocker bugs) and hardware (e.g., NAND flash blocks, cache blocks, logic blocks); however, this makes it difficult to \textit{precisely} discern what a ``block'' is.
In this work, we make little (\textit{negative}) effort to disambiguate these terms and propose our own set of overloaded terms to increase the complexity of this paper.

To inspire new students to join their research groups, professors often hang posters or other publications along the walls adjacent to their offices. 
Regrettably, Saugata does not have any posters directly covering his door, leaving prime real-estate.\footnote{The average price of land in Urbana, IL is roughly \$4.3/sq. ft.}
Therefore, this underutilized space in the Siebel Center for Computer Science offers substantial opportunities for renovations.
To alleviate this concern, we propose Block-SSD. 
Block-SSD takes a basic block, formed out of a page, and physically combines these blocks into larger blocks.
Those blocks are then formed into a larger door block, which cover most of the professor's (e.g., Saugata Ghose~\cite{GhoseWeb}) door.
To our knowledge, we are the first to design a \textit{block-based blocking \underline{\textbf{S}}abotaging \underline{\textbf{S}}augata's \underline{\textbf{D}}oor} (SSD) architecture.

\section{Background}
\paragraph{Blocks.}
Blocks are sometimes but not always formed out of smaller units.
For example, a NAND flash block is made out of smaller units called \textit{pages}.
However, this is not always the case, as operating systems often use pages as virtual blocks.
To make more confusing, a cache block is often a subset of a total data page.
Luckily, there are no paging statements (as opposed to blocking statements in Verilog) because what would they even do?\footnote{Paging statements do exist in limited applications (e.g., Databases)}

For simplicity, and no generality, our work uses a hierarchy of blocking, which mirrors (but still blocks) the NAND flash block paradigm.
Our most \textit{basic block} is formed out of a basic \textit{page} (page for simplicity). 
We connect these horizontally to form \textit{blocker lines}; similarly, along the vertical dimension, we call them \textit{blocking lines}. 
To distinguish Block-SSD from prior work, rather than name the larger block a \textit{superblock}, we instead use the term, \textit{door block}.

\paragraph{SSD Architecture.}
Prior work~\cite{Ghose2023} implemented an SSD architecture in which the students \textbf{kindly} began renovations on Saugata's door.
This was a \textit{complete} misunderstanding\textsuperscript{\texttrademark}, as renovations were not set to begin.\footnote{Saugata was still invoiced for the cost; the balance remains outstanding.}
Unfortunately, although this work successfully implemented a blocking SSD, it was not a block-based blocking SSD.
Therefore, our work SSD differs, in that we are the first block-based blocking SSD.

\section{Key-Idea: A Blocking SSD for Blocking Entry}
When designing an SSD architecture, we consider three main metrics.
First, \textbf{humor}: notably, will Saugata laugh or (threaten to) fire everyone involved.\footnote{Submissions are usually 0-way blind}
Second, \textbf{cost}: Is the absolute dollar amount minimally intrusive to Ryan's funding.
Third, \textbf{visibility}: 'nuff said.
We believe that Block-SSD satisfies all three constraints, as it is funny (\textbf{Laugh}), relatively inexpensive ($<$ 0.1\% of total operating budget), and is visible to anyone who walks by the office in Siebel 4120 (and more \smiley{}).

\section{Methodology \& Results}
We do not elect to use MQSim~\cite{Tavakkol2018}, a widely-used SSD simulator, as Saugata aided in its development, presenting a conflict of interest.
Thus, we forgo simulation and directly fabricate an \SI{88}{\centi\meter}$\times$\SI{211}{\centi\meter} door block.
We fold 280$\pm$20 pages of 8.5''$\times$11'' paper into basic blocks using a previously proposed technique on YouTube~\cite{Block}.
Block-SSD includes a convenient \SI{20}{\centi\meter}$\times$\SI{27}{\centi\meter} non-blocking space for door handles.

\begin{figure}[h]
    \vspace{-15pt}%
    \captionsetup[subfloat]{captionskip=0pt}%
    \subfloat[5$\times$5 Non-blocking tower]%
    {\includegraphics[width=0.24\textwidth]{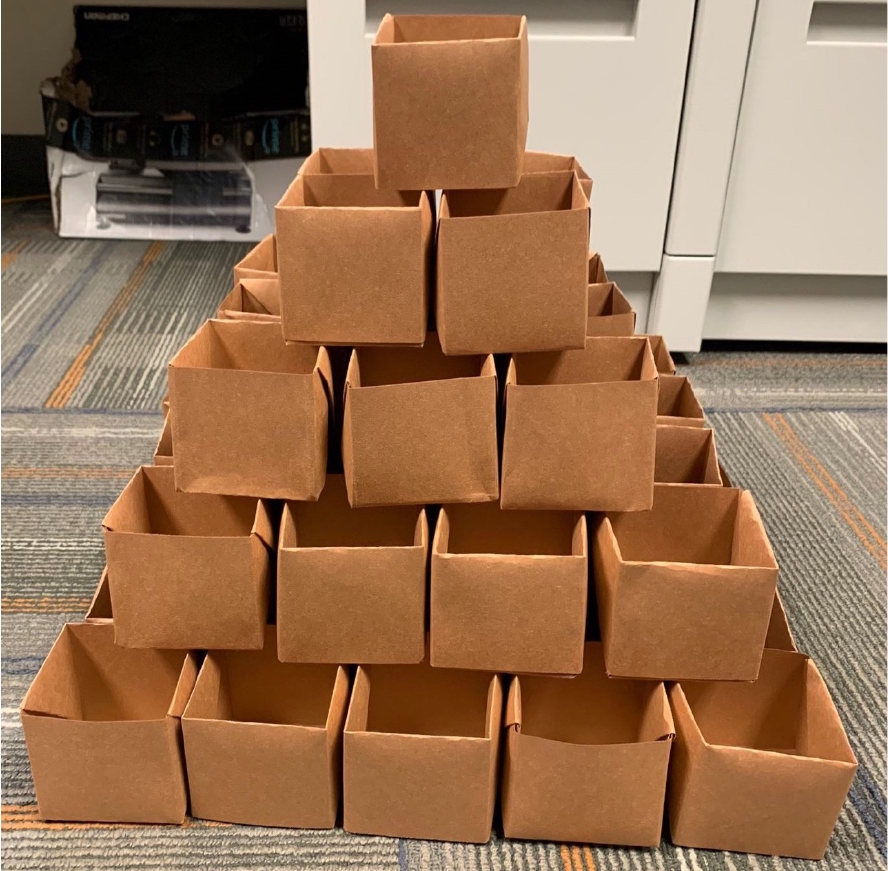}%
    \label{fig:tower5}%
    }%
    \hfill%
   \subfloat[9$\times$9 Non-blocking tower]%
    {\includegraphics[width=0.24\textwidth]{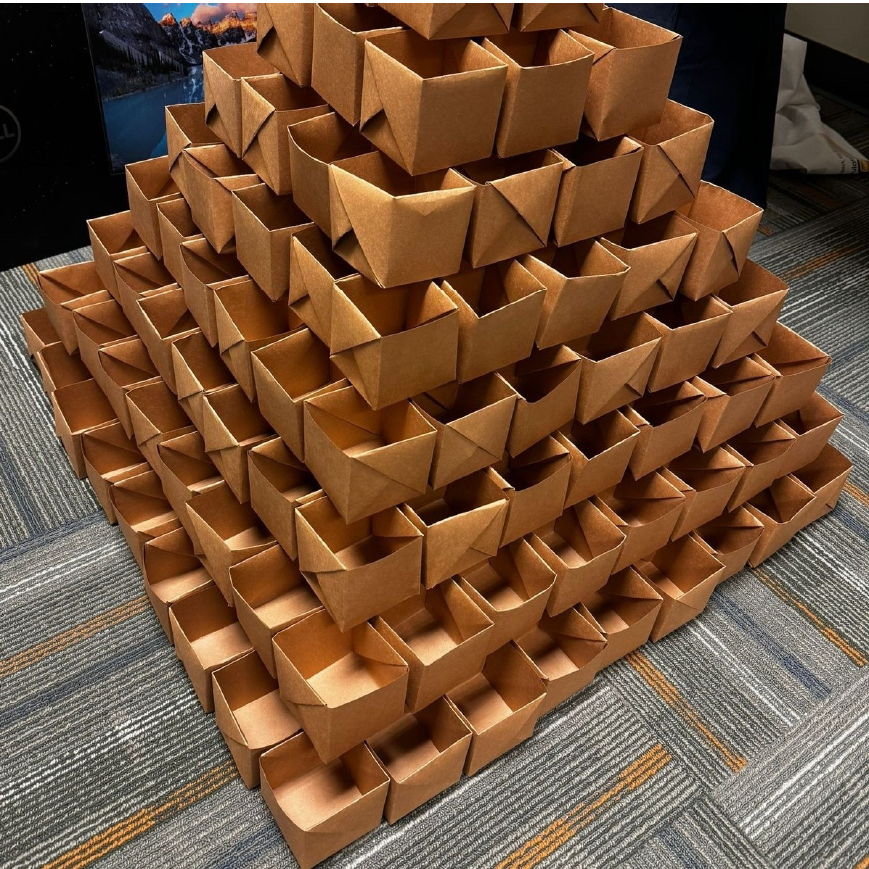}%
    \label{fig:tower9}%
    }%
    \hfill%
    \vspace{-5pt}
    \caption{Work-in-progress non-blocking towers}
    \vspace{-15pt}
    \label{fig:towers}
\end{figure}

\paragraph{Results.} We first verify the integrity of our idea by designing two non-blocking towers.
Figure~\ref{fig:tower5} shows a non-blocking 5$\times$5 non-blocking tower, designed, implemented, and verified by Yiqiu.
Similarly, Figure~\ref{fig:tower9} shows a non-blocking 9$\times$9 non-blocking tower produced by Arjun.
From the figures, we make one surprising observation: \textit{although they are non-blocking towers, they can not be seen through}.
Therefore, they may actually be considered blocking towers...even though they fall short of a true blocking SSD architecture.
\ifcameraready
End Result~\cite{Ghose2024}.
\fi

\section{Conclusion}
In this work, we present Block-SSD, the April 1, 2024, SSD architecture.
Based on our analysis, we appreciate our advisor and believe the feeling is mutual.\footnote{Ryan is the exception. Interpretation left as an exercise for the reader.}
We hope that this line of work can inspire future students (\textbf{WITH THEIR ADVISORS (rough) APPROVAL}) in new blocking architectures.\footnote{The ARCANA group holds no responsibility for the firing of any student.}

\footnotesize{\interlinepenalty=10000
\setstretch{1}
\bibliographystyle{IEEEtranS}
\bibliography{refs.bib}
}

\end{document}